\documentclass[manuscript]{aastex631}

\usepackage{amsmath}

\newcommand{\2}{$_2$}

\shorttitle{}
\shortauthors{}

\begin{document}

\title{Prospects for water vapor detection in the atmospheres of temperate and arid rocky exoplanets around M-dwarf stars}

\correspondingauthor{Feng Ding}
\email{fengding@g.harvard.edu}

\author[0000-0001-7758-4110]{Feng Ding}
\affiliation{School of Engineering and Applied Sciences, Harvard University, Cambridge, MA 02138, USA}

\author[0000-0003-1127-8334]{Robin D. Wordsworth}
\affiliation{School of Engineering and Applied Sciences, Harvard University, Cambridge, MA 02138, USA}
\affiliation{Department of Earth and Planetary Sciences, Harvard University, Cambridge, MA 02138, USA}

\begin{abstract}
Detection of water vapor in the atmosphere of temperate rocky exoplanets would be a major milestone on the path towards characterization of exoplanet habitability. Past modeling work has shown that cloud formation may prevent the detection of water vapor on Earth-like planets with surface oceans using the James Webb Space Telescope (JWST). Here we analyze the potential for atmospheric detection of H\2O on an a different class of targets: arid planets. Using transit spectrum simulations, we show that atmospheric H\2O may be easier to be detected on arid planets with cold-trapped ice deposits on the surface, because such planets will not possess thick H\2O cloud decks that limit the transit depth of spectral features. However, additional factors such as band overlap with {CO\2} and other gases, extinction by mineral dust, {overlap of stellar and planetary H\2O lines,} and the ultimate noise floor obtainable by JWST still pose important challenges. For this reason, combination of space-  and  ground-based spectroscopic observations will be essential for reliable detection of H\2O on rocky exoplanets in the future.
\end{abstract}


\section{Introduction} \label{sec:intro}
Water vapor has now been detected in the atmospheres of {many} gas-giant exoplanets \citep{fortney2021hotjupiters}, and in some cases abundance constraints have been derived from the transmission spectra (e.g., \citealt{madhusuhan2014h2o,barstow2017retrieval,pinhas2019retrieval}). An important goal for future large ground- and space-based telescopes is to detect water vapor molecules in the secondary atmospheres of rocky exoplanets. In particular, detection of Earth-like water vapor levels on rocky planets orbiting within the habitable zones of their host stars would suggest the existence of a surface liquid water reservoir -- a necessary condition for the survival of Earth-like life \citep{Kasting1993habitable,kopparapu2013clima}. Several recent studies have simulated the spectral features of Earth-like aqua-planets (with global surface entirely covered in oceans) around low-mass stars using three dimensional general circulation models (3D GCMs). These studies showed that ice cloud particles suspended in the upper atmosphere resulting from tropospheric deep convection and the large-scale circulation will mute {H\2O} features in the transmission spectra of future space-based observations, e.g., the James Webb Space Telescope (JWST), likely rendering water vapor undetectable \citep{fauchez2019trappist1,komacek2020cloud,Pidhorodetska2020trappist1e,suissa2020dim}.

\begin{figure}[ht]
  \centering
  \includegraphics[width=\columnwidth]{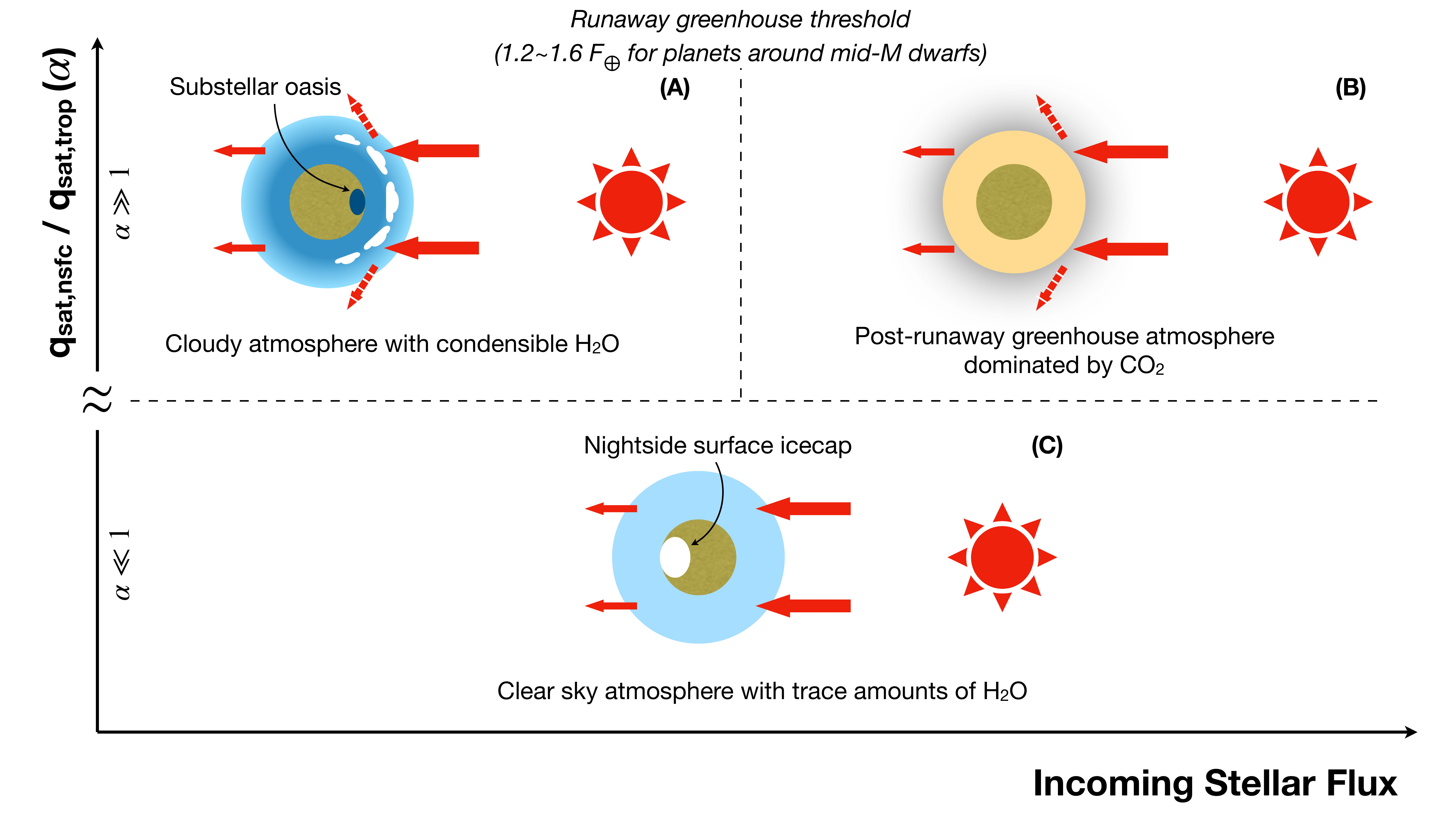}
  \caption{Schematic representation of the multiple moist climate equilibrium states on synchronously rotating arid planets when the surface water is trapped by the substellar tropopause as a substellar oasis (a), entirely evaporated into the atmosphere in a post-runaway greenhouse climate state (b), and is trapped by the nightside surface as ice caps (c). The vertical axis represents the cold trapping strength between the nightside surface and substellar tropopause. The hydrological cycle is dominated by the cold trap on the nightside surface when $\alpha \ll 1$ and at the substellar tropopause when $\alpha \gg 1$.  }\label{fig:schematic}
\end{figure}

However, the detectability of water vapor on rocky planets with limited surface water reservoirs (hereafter referred to as arid planets) has not yet been studied in detail. Many M-dwarf rocky planets may be water-poor as a 
result of their host stars' prolonged pre-main-sequence phase, which may cause extensive early water loss \citep{ramirez2014premain,luger2015waterloss,tian2015waterloss}. Recent GCM studies have revealed multiple moist climate equilibrium states on such arid planets (e.g., \citealt{leconte2013bistable,ding2020arid,ding2021lastsat}). Figure~\ref{fig:schematic} illustrates the multiple moist climate states as a result of the cold trapping competition between the substellar tropopause and the nightside surface, which was discussed in more detail in \citet{ding2020arid} and \citet{ding2021lastsat}. When the cold trap at the substellar tropopause dominates the hydrological cycle on arid planets, the surface water reservoir can be either trapped by the substellar tropopause as a substellar oasis, or entirely evaporated into the atmosphere \citep{ding2020arid}. The outcome depends on whether the incoming stellar radiation exceeds the runaway greenhouse threshold ($1.2 \sim 1.6 F_\oplus$ for planets around mid-M dwarfs, where $F_\oplus$ is the present-day Earth's incoming stellar flux, see \citealt{yang2014innerrotation,kopparapu2016inneredge}). 

When the cold trap on the nightside surface dominates the hydrological cycle on arid planets, climate models predict that the surface water reservoir will be trapped on the nightside as ice caps. In this situation, the tropopause cold trap becomes ineffective and hence the atmospheric water vapor is in vapor-ice phase equilibrium with the surface ice layer and well-mixed in the atmosphere \citep{ding2020arid,ding2021lastsat}. 
This particular climate state with only nightside surface ice deposits and no dayside condensation clouds can exist under much higher stellar radiation beyond the runaway greenhouse threshold, because the extremely under-saturated dayside atmosphere behaves as an effective radiator fin  \citep{pierrehumbert_thermostats_1995,abe2011dry,zsom2013innerdry,kodama2018dry,ding2020arid} and the optically thin atmosphere redistributes heat inefficiently between the dayside and nightside \citep{leconte2013bistable,wordsworth2015heat,koll2016heat}. A partial analogy to this climate state can be seen in the solar system. Water ice is present near Mercury's north pole, despite the fact that this planet receives an averaged stellar flux of $\sim 6.7 F_\oplus$ \citep{lawrence2013mercury}. Some nearby transiting temperate rocky planets receiving stellar fluxes between $2F_\oplus$ and $6F_\oplus$ have been confirmed recently, such as L~98-59d, LHS~1140c, LTT~1445Ab, TRAPPIST-1b and c \footnote{These planets are in multiple planetary systems and some of them may not be synchronously rotating. But water ice can still exist in polar regions on asynchronously rotating planets if they are cold enough, as on Mercury and the Moon.}. Whether these planets have a thick Venus-like post-runaway greenhouse or an $\mathcal{O}(1)$ bar atmosphere can be distinguished by near future observations of thermal phase curve or eclipse photometry  \citet{selsis2011phasecurve,koll2015phasecurve,koll2019ireclipse}. A near-infrared (NIR) phase curve has been observed for LHS~3844b, a hotter rocky exoplanet receiving stellar flux of $70\,F_\oplus$. 
The symmetric shape and large amplitude of the phase curve rule out thick atmospheres (denser than 10\,bar) on LHS~3844b \citep{kreidberg2019lhs3844b}.
Therefore, transit spectroscopic observations combined with thermal phase curve observations will be an  ideal tool to characterize atmospheres and identify whether surface ice deposits can survive in cold-trap regions on temperate rocky exoplanets. The aim of this letter is to investigate the detectability of water vapor on planets in this climate state and compare with previously simulated results for habitable aqua-planets.

\section{Method} \label{sec:method}
\subsection{GCM setup}

We simulate an Earth-sized synchronously rotating planet orbiting an M5V-type star with an effective temperature of 3060\,K. We use the idealized GCM developed in \citet{ding2019fmspcm} to simulate climates on temperate arid exoplanets with nightside ice deposits. Our GCM uses a line-by-line approach to calculate radiative transfer, which maintains both flexibility and accuracy for simulating diverse planetary atmospheres. We assume the surface is bare rock with an albedo of 0.2. The atmosphere is made of 1\,bar N\2 and variable amounts of H\2O. The effects of varying atmospheric composition beyond this idealized Earth-like case, and in particular the effects of adding CO\2 to the atmosphere, are discussed in subsequent sections. We run a series of simulations with various incoming stellar fluxes to determine the critical flux under which the initial H\2O can condense on the nightside surface and form ice layers. 
For each given incoming stellar flux $F$, the model is integrated for 3000\,days. The corresponding orbital period $P$ of the planet is calculated self-consistently using Kepler's third law \citep{wordsworth2015heat,kopparapu2016inneredge,haqq2018circulation} as
\begin{equation}
P = 365\, \mathrm{ days} \left( \frac{L}{L_\odot} \right)^{3/4} \left( \frac{M}{M_\odot} \right)^{-1/2} \left( \frac{F}{F_\oplus} \right)^{-3/4}
\end{equation}
where $L/L_\odot = 0.003$ is the luminosity of the host star scaled by the luminosity of the Sun and $M/M_\odot = 0.16$ is the mass of the host star in solar mass units \citep{pecaut2013stars}. 

\subsection{Simulation of transmission spectra }

We use the Planetary Spectrum Generator\footnote{\url{https://psg.gsfc.nasa.gov}} (PSG, \citealt{Villanueva2018psg}) to simulate the transmission spectra of transiting planets with nightside ice deposits. In this study we make a conservative estimate by assuming that the planetary system is 15 parsec away, further than the confirmed nearby temperate rocky exoplanets mentioned earlier. The parameters of the host star and the simulated planet are the same as those used in the GCM simulation.
We use PSG to simulate the near-infrared transmission spectra between 0.6 and 5.3 $\mu$m with a resolving power $R=50$, which is relevant for the JWST Near-Infrared Spectrograph (NIRSpec)/PRISM instrument. We follow {\citet{louie2018transit} and} \citet{yaeger2019trappist} to estimate the total expected signal-to-noise ratio (S/N) as the quadrature sum of individual S/N from each spectral element when detecting water vapor features in the transmission spectra. We assume the out-of-transit observation time the same as the in-transit time and the noise level dominated by photon noise statistics. For detection of molecular features, we assume a signal-to-noise ratio of 5 is required.

\section{The Dependence of Transmission Spectra on Atmospheric Parameters} \label{sec:spectra}

\begin{figure}[ht]
  \centering
  \includegraphics[width=\columnwidth]{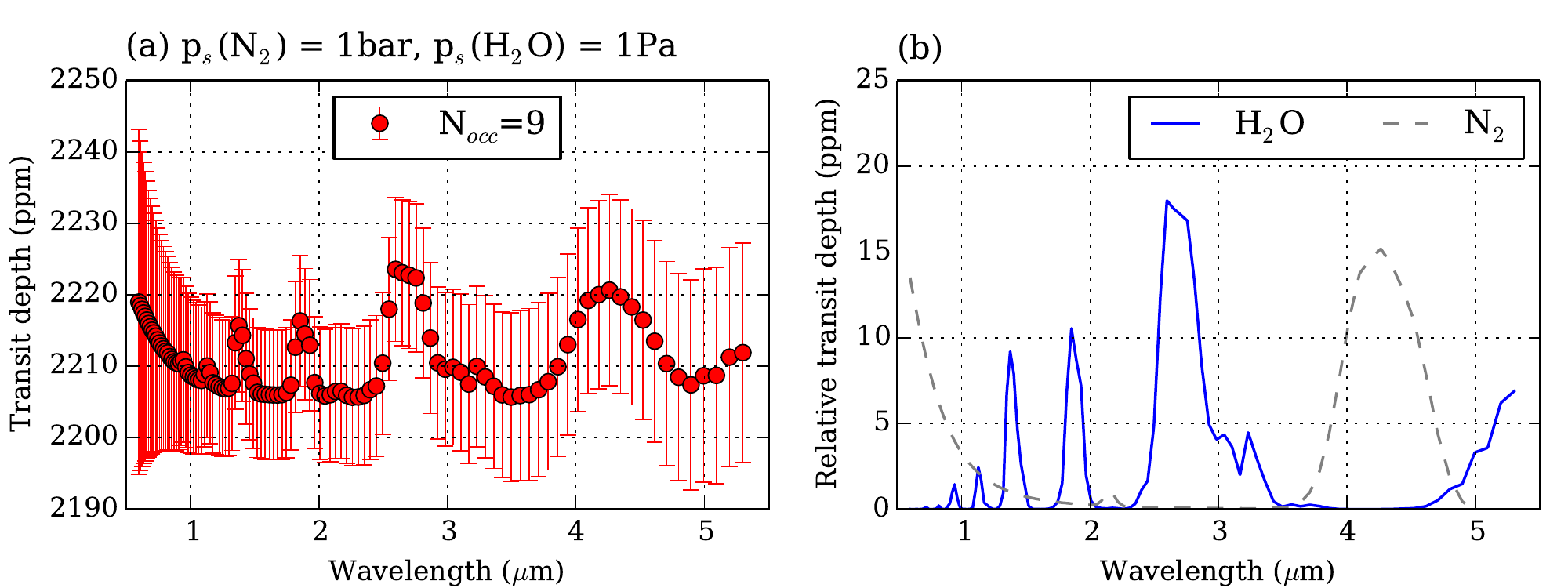}
  \caption{(a) Simulated JWST/NIRSpec Prism transmission spectra of a synchronously rotating planet with the nightside ice deposits. The atmosphere is made of 1\,bar N\2 and 10\,ppmv H\2O. $N_{occ}=9$ is the number of transits needed to detect water features at an S/N=5. Red points with 1$\sigma$ error bars represent the simulated spectra from 9 transits.  (b)  The individual molecular absorption features for N\2 (gray dash) and H\2O (blue solid). }\label{fig:snrclear}
\end{figure}

We first estimate the climate on temperate arid exoplanets with nightside ice deposits by performing a series of GCM simulations similar to the atmospheric collapse simulations in \citet{wordsworth2015heat}. We find that when the incoming stellar flux decreases to $6000\,\mathrm{W\,m^{-2}}$($4.4F_\oplus, P = 3.86$ days) water vapor starts to condense on the nightside surface for an atmosphere made of 1\,bar N\2 with 10\,ppmv H\2O. This water vapor mixing ratio is close to the value in the upper stratosphere of present-day Earth \citep{park2021stratospherewv} and the corresponding temperature of the nightside surface cold trap is 213\,K.
We then use PSG to simulate the JWST/NIRSpec Prism transmission spectra of the clear-sky atmosphere as water vapor starts to condense on the surface and show it in Figure~\ref{fig:snrclear}.  Only 9 transits are required to detect this small quantity of H\2O in the clear-sky atmosphere with total expected S/N=5 (the duration time of one single transit is 3879\,s). The absorption feature by water vapor at 2.7\,$\mu$m is most observable and the relative transit depth can reach 17\,ppmv. In comparison, the relative transit depth at 2.7\,$\mu$m is less than 7\,ppm for aqua-planets around M5V-type stars because of the muting of spectral features by water ice clouds in the upper atmosphere \citep{suissa2020dim}. 

\begin{figure}[ht]
  \centering
  \includegraphics[width=0.6\columnwidth]{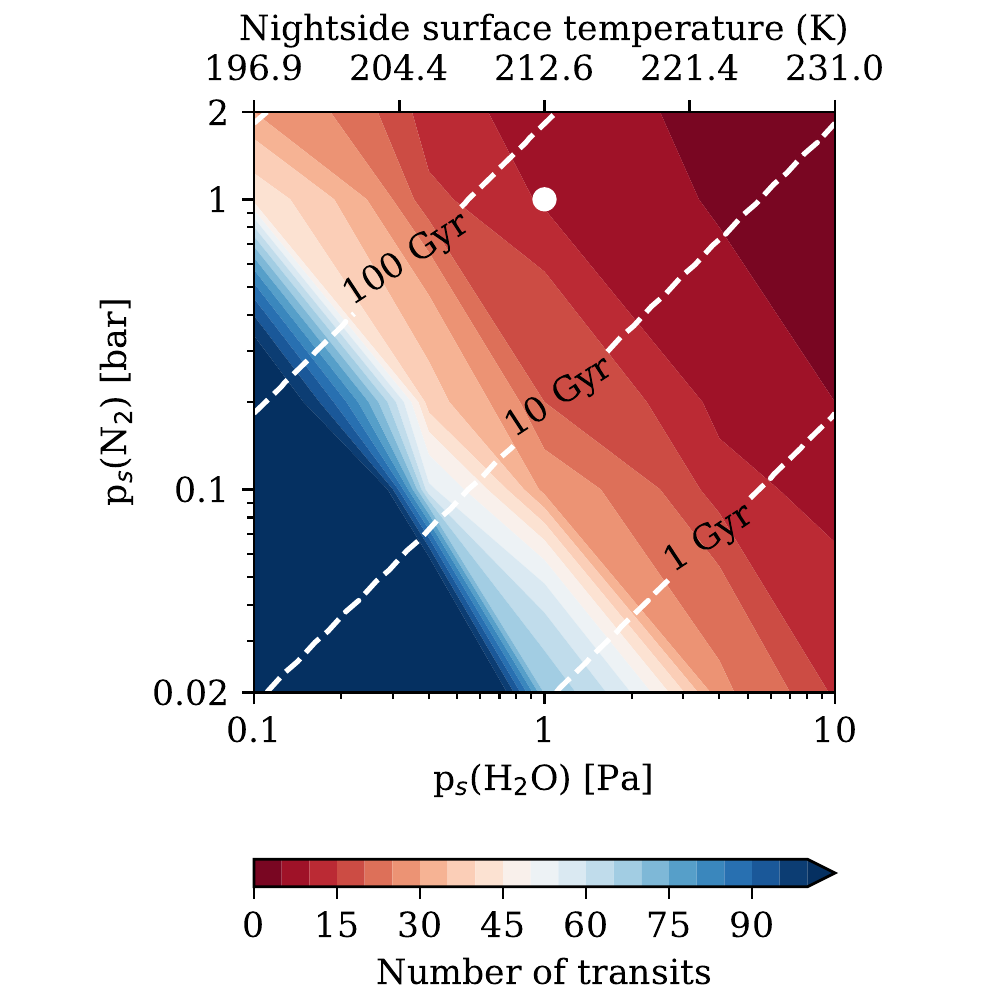}
  \caption{Required number of transits to detect water vapor features in the transmission spectra as a function of surface partial pressure of N\2 and H\2O. The respective nightside surface ice temperatures in ice-vapor phase equilibrium with atmospheric water vapor are labeled in the upper x-axis. The white dashed lines show the lifetime of surface ice deposits estimated by the diffusion-limited escape flux. The surface ice layer is assumed to be 1\,km thick and cover 10\% of the global surface area. The climate state used to calculate the transmission spectrum in Figure~\ref{fig:snrclear} is marked by the white dot.}\label{fig:n2wv}
\end{figure}

Next, we use PSG to explore how atmospheric N\2 and H\2O inventories can affect the detectability of water vapor features in the transmission spectra by varying the surface partial pressure of N\2 from 0.02\,bar to 2\,bar and H\2O from 0.1\,Pa to 10\,Pa. Here we simply assume the same orbital radius and period as used in Figure~\ref{fig:snrclear}. This assumption has negligible effect on the transmission spectra calculation, but fixes the transit duration time and therefore the photon-noise statistics. The required  number  of  transits  to  detect  water  vapor  features is shown in Figure~\ref{fig:n2wv}. The simulation results suggest water vapor features are observable in the transmission spectra, but only over a relatively narrow parameter range. For instance, in order to detect water vapor on a temperate arid planet with nightside ice deposits within less than 30 transits, the temperature of the nightside surface cold trap should be above 200\,K if the background atmosphere is 1\,bar N\2. 

For thinner background atmospheres, this required temperature is higher because collisions among air molecules are less frequent and higher H\2O {concentrations} are therefore required {to be detected}. On temperate arid planets, the surface temperature of the nightside cold trap depends on not only the incoming stellar flux but also many planetary parameters, such as planetary rotation rate and atmospheric composition (see also Section~\ref{sec:conclusion}).  Recent work has suggested that turbulent mixing near the nightside surface should be stronger than simulated by GCMs \citep{joshi2020boundary}. So further studies combined with eddy-resolving models and large-scale GCMs are required to fully understand the nightside near surface conditions that are crucial for surface cold trapping.

An important factor determining the lifetime of surface ice deposits is atmospheric escape. In the parameter space we explore here, water vapor is the minor species in the atmosphere. In this circumstance, the atmospheric escape flux would most likely be limited by diffusion through the homopause. Here we estimate the lifetime of surface ice by using the diffusion-limited H\2 escape flux (measured by number of escape particles per unit surface area of the planet per unit time) following \citet{abe2011dry}
\begin{equation}
    F_{dl} = f_{str}(\mathrm{H_2})\ b\ \frac{(m_\mathrm{N_2} - m_\mathrm{H_2})g}{k_B T_{str}}
\end{equation}
where $f_{str}(\mathrm{H_2})$ is the mixing ratio of hydrogen in all forms at the homopause, $T_{str}\sim300\,$K is the homopause temperature, $m_\mathrm{N_2}$ and $m_\mathrm{H_2}$ are the particle mass of N\2 and H\2, and $b=1.9\times10^{21} (T/300)^{0.75}\,\mathrm{m^{-1}s^{-1}}$ is the binary diffusion coefficient of H\2 in N\2. If the diffusion species is H instead of H\2 above the homopause, the escape flux is increased by a factor of 2. Assuming the surface ice layer to be 1\,km thick and cover 10\% of the global surface area, we find the lifetime of the ice deposits for most parameters we consider here is greater than 1\,Gyr. As a comparison, without the supply of water vapor from the surface ice deposits, the e-folding lifetime of atmospheric water vapor is shorter than 70\,kyr. Combining the calculations of transmission spectra and surface ice lifetime, it is possible to detect water vapor features on temperate arid planets with nightside ice deposits when the background atmospheric pressure is above 0.05\,bar and the nightside surface temperature is above 200\,K. 

\begin{figure}[ht]
  \centering
  \includegraphics[width=0.6\columnwidth]{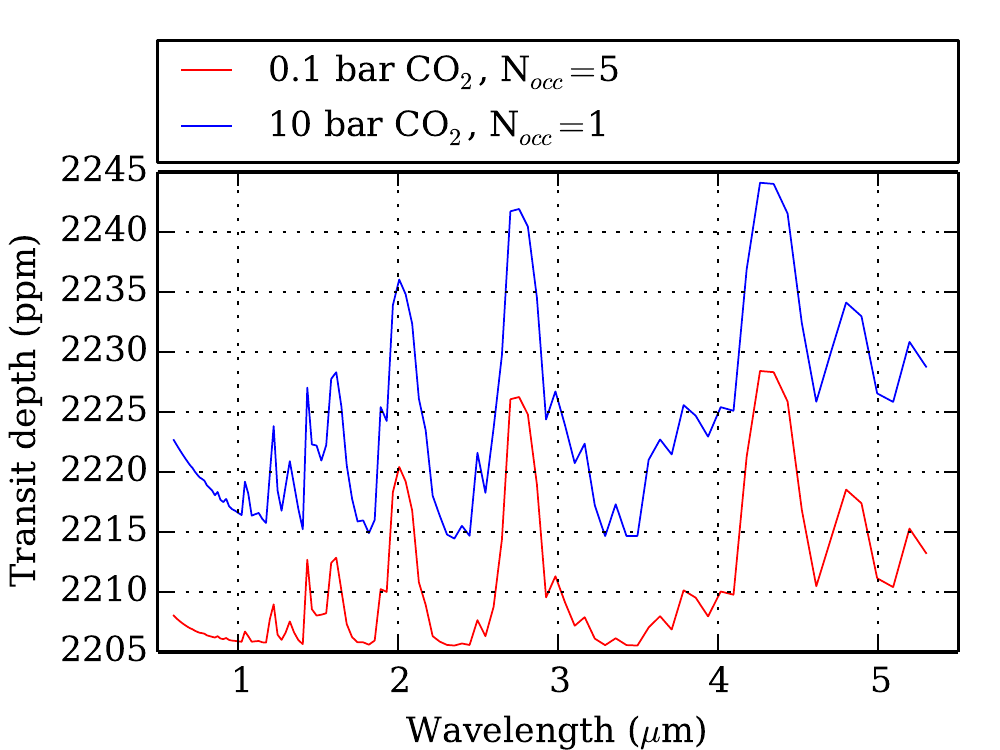}
  \caption{ Simulated JWST/NIRSpec Prism transmission spectra of a synchronously rotating planet with 10\,bar CO\2 (blue) and 0.1\,bar CO\2 (red). $N_{occ}$ is the number of transits needed to detect CO\2 features at an S/N=5. For the 0.1\,bar CO\2 case, atmospheric CO\2 can condense on the nightside and form CO\2 ice deposits as on present-day Mars. Both cases assume the clear-sky condition.}\label{fig:co2}
\end{figure}

Finally, it is worth noting that water vapor is not the only volatile species that can condense on the nightside on temperate rocky exoplanets. There are many other candidates given right conditions. Another common condensing species is CO\2. \citet{wordsworth2015heat} investigated the relation between the atmospheric CO\2 pressure and the critical incoming stellar flux to trigger nightside surface condensation (ranging from $0.2\,F_\oplus$ to $3\,F_\oplus$), focusing on slowly rotating planets around bright M-dwarfs. Using our GCM, we similarly find that for a planet with 0.1\,bar CO\2 atmosphere around a bright M-dwarf the  critical incoming stellar flux that triggers the nightside surface condensation is $\sim 1\,F_\oplus$ \citep{ding2019fmspcm}. When the host star is M5V-type, we find the critical flux increases to $4000\,\mathrm{W\,m^{-2}} (2.9\,F_\oplus)$. The corresponding orbital radius and period are 0.032\,AU and 5.23\,days, respectively. Then we use PSG to simulate the transmission spectrum of such a transiting planet, which is Mars-like with a CO\2-dominated atmosphere but CO\2 ice deposits forming on the nightside rather than around the polar regions. The climate resembles the moist climate described in Figure~\ref{fig:schematic}c but with a condensing CO\2 atmosphere.
We compare the result with the case of a dense 10\,bar CO\2 atmosphere in Figure~\ref{fig:co2}. 

Both cases show stronger NIR absorption features than the 10\,ppmv water vapor case in Figure~\ref{fig:snrclear} so that fewer transits are required to detect CO\2 features in the transmission spectra. Another major difference is that atmospheric refraction plays an important role in dense CO\2 atmospheres. For temperate planets with CO\2-dominated atmosphere around M5V-type stars, PSG simulation results show that only the top 0.7\,bar of atmosphere can be probed. This 0.7\,bar layer mimics a surface (e.g., \citealt{hui2002refraction,misra2014refraction,betremieux2015refraction}) and raises the baseline of the transmission spectrum of the dense CO\2 atmosphere. So both cases have similar relative transit depths and can be further distinguished by observation of the thermal phase curve or eclipse photometry.

\section{Conclusion and Discussions} \label{sec:conclusion}

Temperate rocky exoplanets inside the inner edge of conventional habitable zone can store substantial amount of remaining water ice on the nightside surface if the atmosphere is optically thin in the thermal infrared and redistributes heat inefficiently. Our simulation results suggest that strong water vapor features in transmission spectra of those climate systems can potentially be detected for surface pressures higher than 0.05\,bar and nightside surface temperature higher than 200\,K. NIR transit spectroscopy combined with thermal phase curve or eclipse photometry by JWST will be an ideal tool to identify the climate state of temperate rocky exoplanets covered by remnant ice deposits on the nightside. Phase curve observation has been proved to be very effective to distinguish whether the atmosphere is optically thin or thick on a hot rocky exoplanet, LHS 3844b \citep{kreidberg2019lhs3844b}. If water vapor was detected in the transit transmission spectrum of a temperate rocky exoplanet with an optically thin atmosphere, it is very likely that the water vapor would be in phase equilibrium with remnant nightside ice deposits, because the lifetime of water vapor in such atmosphere is usually shorter than 0.1\,Myr without a surface supply.


There are several important factors that could complicate the analysis presented here. First, our assumption of an atmosphere only made of N\2 and H\2O is very simplified. In reality, other gaseous species in the atmosphere are expected, such as CO\2, O\2, O$_3$, H\2, CO and CH$_4$, depending on the redox evolution of the planet and other effects. These gases can provide additional thermal infrared opacity and raise the surface temperature on the nightside. If the IR opacity is not high enough to trigger a runaway greenhouse effect and sublimate the entire nightside ice layer, the existence of these gases in fact increases the atmospheric water vapor concentration and makes water vapor more detectable. Among these gases, CO\2 is most important because its absorption bands overlap with H\2O at 1.8\,$\mu$m and 2.7\,$\mu$m. If only the 1.4\,$\mu$m absorption band is used for the water vapor detection calculation in Figure~\ref{fig:snrclear}, then at least 80 transits would be required. 

Mineral dust is another complicating factor. Dust could be lifted by convective plumes around the substellar area on arid planets \citep{boutle2020dust}. Dust particles transported  by the large-scale circulation to the night hemisphere can increase the thermal infrared opacity and warm the nightside surface. Near the terminator, dust particles suspended in the atmosphere can obscure the water vapor features in transmission spectra, in a similar way to ice clouds on aqua-planets. Understanding the impacts of mineral dust on the climates of temperate rocky exoplanets with nightside ice deposits will require climate modeling that incorporates dust cycles.

{Stellar models suggest the spectra of M-dwarfs are likely to be enriched in water absorption lines, especially for late M-dwarfs \citep{deming2017RLB,reiners2018mdwarf}. \citet{deming2017RLB} showed including the overlap of stellar M-dwarf and planetary water lines at high spectral resolution can reduce the modeled transmission transit depth by $\sim10\%$, referred to as spectral resolution-linked bias (RLB) in their study of TRAPPIST-1b. This RLB effect is not taken into account in the PSG simulation, which uses pre-computed correlated-k tables to calculate radiative transfer for resolving powers less than 5000 \citep{Villanueva2018psg}. The importance of water line overlaps between the spectra of temperate arid exoplanets and mid M-dwarfs  should be  evaluated by future line-by-line analyses. } 

Finally, the noise floor induced by the telescope in its environment is a key factor when detecting molecular absorption features in transmission spectra, especially with low spectral depths \citep{greene2016noisefloor,suissa2020dim}. All calculations in this paper impose no absolute noise floor on observations. In our simulations, the absorption feature at 2.7\,$\mu$m is most observable (Figure~\ref{fig:snrclear}b) and stronger than those in simulations for habitable aqua-planets around the same type of stars. But  a noise floor as low as 3\,ppm is still required otherwise the photon noise will reach the noise floor and any signal from the star is drown out. If the noise floor in the near-infrared is in the range of 10-20\,ppm, as estimated by \citet{greene2016noisefloor}, none of the simulation results for temperate planets around M5V-type stars in Figure~\ref{fig:n2wv} can be detectable and only temperate planets with nightside ice deposits around late M-dwarfs might be detectable with JWST. 
As a result, synergies between future space-based spectroscopy 
and ground-based spectroscopy (such as the Extremely Large Telescope (ELT) using cross-correlation techniques), building on past successes for hot Jupiters \citep[e.g., ][]{brogi2017lowandhigh,brogi2019crosscorre}, may be essential in order to overcome these difficulties associated with absorption band overlap with CO\2 and the noise floor.

\begin{acknowledgments}
We thank the referee for thoughtful comments that improved the manuscript. R.W. acknowledges funding support from NASA/VPL grant UWSC10439. The GCM simulations in this paper were run on the FASRC Cannon cluster supported by the FAS Division of Science Research Computing Group at Harvard University. 
\end{acknowledgments}

\bibliographystyle{aasjournal}

\end{document}